\documentclass[onecolumn,superscriptaddress]{revtex4}
\usepackage{amssymb}
\usepackage{amsmath}

\begin{document}

\title{Testing nonlinear electrodynamics in waveguides:
the effect of magnetostatic fields on the transmitted power}
\author{Rafael Ferraro}
\email{ferraro@iafe.uba.ar}
\thanks{Member of Carrera del Investigador Cient\'{\i}fico (CONICET,
Argentina)} \affiliation{Instituto de Astronom\'\i a y F\'\i sica
del Espacio, Casilla de Correo 67, Sucursal 28, 1428 Buenos Aires,
Argentina} \affiliation{Departamento de F\'\i sica, Facultad de
Ciencias Exactas y Naturales, Universidad de Buenos Aires, Ciudad
Universitaria, Pabell\'on I, 1428 Buenos Aires, Argentina.}

\begin{abstract}
In Born-Infeld theory and other nonlinear electrodynamics, the
presence of a magnetostatic field modifies the dispersion relation
and the energy velocity of waves propagating in a hollow
waveguide. As a consequence, the transmitted power along a
waveguide suffers slight changes when a magnetostatic field is
switched on and off. This tiny effect could be better tested by
operating the waveguide at a frequency close to the cutoff
frequency.
\end{abstract}


\pacs{03.50.Kk, 03.65.Pm, 42.82.Et}

\keywords{waveguides, non-linear electrodynamics, Born-Infeld}

\maketitle

\section{Introduction}
Born-Infeld electrodynamics \cite{BI,BI2,BI3,BI4} was created with
the aim of healing the divergence of the point-like charge
self-energy, but nowadays became the object of great interest
because of its relation with the low energy dynamics of strings
and branes \cite{Frad, Abou,Leigh,Met,Tsey,Tsey2}. As a nonlinear
extension of Maxwell's theory, Born-Infeld electrodynamics has the
quality of being the sole extension free of birefringence (i.e.,
the wave-vector of a propagating perturbation results to be
univocally determined) \cite{Pleb,Deser,Boillat,Novello}. Few
exact solutions of Born-Infeld theory are known, but some issues
concerning the wave propagation have been thoroughly studied.
Namely, it is known that the propagation velocity is $c$\ for free
perturbations traveling in vacuum, but it is lower than $c$ if an
external field is present \cite{Pleb,5pleb,Boillat}; in other
words, the external fields affect the dispersion relation. The
dispersion relation is also affected by superposition of free
waves. In fact, although the Born-Infeld free waves are identical
to those of Maxwell's theory, the Born-Infeld nonlinearity causes
interactions among free waves that influence the dispersion
relation. This subject was studied in a previous paper by solving
the Born-Infeld field equations in a cavity where stationary waves
are formed \cite{ferr}. Since a stationary wave is the
superposition of free waves bouncing backwards and forwards, the
interaction between them leads the wave amplitude to enter the
dispersion relation in a waveguide. In this paper we will show
that even a magnetostatic field affect the dispersion relation of
waves propagating in a hollow waveguide by increasing the energy
velocity. As a consequence, the transmitted power along the
waveguide could be controlled by means of magnetostatic fields.
Although this nonlinear electrodynamics effect should be very
tiny, it has a better chance of being tested when the waveguide
works at a frequency close to the cutoff frequency.

\section{Born-Infeld field}
Like in Maxwell's electromagnetism, Born-Infeld field $F_{\mu \nu
}$ is derived from a potential: $F_{\mu \nu }=\partial _{\mu
}A_{\nu }-\partial _{\nu }A_{\mu }$ (in geometric language, the
2-form $F$ is exact: $F=dA$). This condition cancels the curl of
the electric field $\mathbf{E}$ and the divergence of the magnetic
field $\mathbf{B}$:
\begin{equation}
\partial _{\nu }F_{\lambda \mu }+\partial _{\mu }F_{\nu \lambda }+\partial
_{\lambda }F_{\mu \nu }\ =\ 0\ ,  \label{eq1}
\end{equation}
i.e., $dF=0$. Born-Infeld field differs from Maxwell field in the dynamic
equations, which are written in terms of the tensor
\begin{equation}
\mathcal{F}_{\mu \nu }=\frac{F_{\mu \nu }-\frac{P}{b^{2}}\ ^{\ast }F_{\mu
\nu }}{\sqrt{1+\frac{2S}{b^{2}}-\frac{P^{2}}{b^{4}}}}  \label{eq2}
\end{equation}
where $S$\ and $P$\ are the scalar and pseudoscalar field
invariants:
\begin{equation}
S=\frac{1}{4}\ F_{\mu \nu }F^{\mu \nu }=\frac{1}{2}\ (|\mathbf{B}|^{2}-|
\mathbf{E}|^{2})  \label{eq3}
\end{equation}
\begin{equation}
P=\frac{1}{4}\ ^{\ast }F_{\mu \nu }F^{\mu \nu }=\mathbf{E}\cdot \mathbf{B}
\label{eq4}
\end{equation}
($^{\ast }F_{\mu \nu }$ is the dual field tensor, i.e. the tensor resulting
from exchanging the roles of $\mathbf{E}$ and ${-\mathbf{B}}$). Born-Infeld
dynamical equations are
\begin{equation}
\partial _{\nu }\mathcal{F}^{\mu \nu }=\ 0\ ,  \label{eq5}
\end{equation}
(i.e., $d\,^{\ast }\! \mathcal{F}=0$) which is obtained from the
Born-Infeld Lagrangian
\begin{equation}
L[A_{\mu }]=-\frac{b^{2}}{4\,\pi
}\;\left(1-\sqrt{1+\frac{2S}{b^{2}}-\frac{P^{2}}{b^{4}}}\right)
\label{eq6}
\end{equation}
The constant $b$ in Eqs.~(\ref{eq2}) and (\ref{eq6}) is a new
universal constant with units of field that controls the scale for
passing from Maxwell's theory to the nonlinear Born-Infeld regime,
in the same way as the light speed $c$ is the velocity scale that
indicates the range of validity of Newtonian mechanics. The
Maxwell Lagrangian and its related dynamical equations are
recovered in the limit $b\rightarrow \infty $, or in regions where
the field is small compared with $b$. Besides, Born-Infeld
solutions having $S = 0 = P$ (``free waves") also solve Maxwell's
equations.

\section{Stationary waves}
For our purposes, we will concentrate in those Born-Infeld waves that can be
written as $F=d[u(t,x)]\wedge dy$, i.e.
\begin{equation}
F\ =\ \frac{\partial u(t,x)}{\partial t}\ dt\wedge dy\ +\ \frac{\partial
u(t,x)}{\partial x}\ dx\wedge dy  \label{eq7}
\end{equation}
where the symbol $\wedge$ is the antisymmetrized tensor product.
Expression (\ref{eq7}) means
\begin{equation}
cE_{y}=F_{t\,y}=\partial u/\partial t=-F_{y\,t\;}  \label{eq8}
\end{equation}
\begin{equation}
B_{z}=-F_{xy}=-\partial u/\partial x=F_{y\,x}\ ,  \label{eq9}
\end{equation}
the rest of the components $F_{\mu \nu }$ being zero. So the
pseudoscalar invariant $P$ vanishes for the proposed solution. The
field (\ref{eq7}) accomplishes Eq.~(\ref{eq1}), since $dF$ is
identically null. We will choose the function $u(t,x)$\ to satisfy
boundary conditions suitable for a rectangular box:
\begin{equation}
E_{y}(t,x=0)\,=\,0\,=\,E_{y}(t,x=d).  \label{eq10}
\end{equation}
By substituting the field (\ref{eq7}) in Eq.~(\ref{eq5}), one
obtains the dynamical equation for $u(t,x)$:
\begin{equation}
\mathcal{BI\;}u(t,x)\equiv \left[ 1+\frac{1\,}{b^{2}}\left(
\frac{\partial u}{\partial x}\right) ^{2}\right] \frac{\partial
^{2}u}{\partial t^{2}}-\frac{2\,}{b^{2}}\frac{\partial u}{\partial
t}\frac{\partial u}{\partial x}\ \frac{\partial ^{2}u}{\partial
t\,\partial x}-c^{2}\left[ 1-\frac{1\,}{c^{2}b^{2}}\left(
\frac{\partial u}{\partial t}\right) ^{2}\right] \frac{\partial
^{2}u}{\partial x^{2}}=0 \label{eq11}
\end{equation}
This is the so called Born-Infeld equation, which can be independently
derived from the scalar field Lagrangian $L[u]\varpropto (1\,-\,b^{-2}\ \eta
^{\mu \nu }\ \partial _{\mu }\,u\ \partial _{\nu }\,u)^{1/2}$.

\section{BI-guided waves in the presence of a magnetostatic field}
In Ref.$\,$\cite{ferr} we have used the scheme of Section III to
study waves propagating in a rectangular waveguide along the
$z$-axis. Although the field (\ref{eq7}) does not depend on $z$
-so the field (\ref{eq7}) would be just a stationary wave bouncing
backwards and forwards between two opposite walls of the guide-,
the propagation along the guide can be introduced by transforming
the field with a Lorentz boost in the $z$-direction. This
procedure would convert the solution (\ref{eq7}) into a transverse
electric propagating mode (notice that a boost along $z$ does not
modify the boundary conditions (\ref{eq10})). In this paper we
will study the nonlinear effects produced by the presence of a
magnetostatic field $B_{{\ell }}$ along the waveguide. Equipped
with the knowledge of the solution for a stationary wave between
parallel plates with $B_{{\ell }}=0$ \cite{ferr}, we prepare the
solution with three free parameters -$\alpha$, $\beta$ and
$\Omega$- that will allow us to fit the Eq.~(\ref{eq11}):
\begin{equation}
u(t,x)=\frac{E}{\kappa }\,\Big[\cos (\Omega t)+\frac{E^{2}}{32\
b^{2}} \,\cos (3\Omega t)\Big] \Big[\sin (\kappa
x)+\frac{E^{2}}{32\ b^{2}}\,\sin (3\kappa x)\Big]-B_{{\ell
}}\Big[x+\alpha \ \frac{E^{2}}{\kappa b^{2}}\ \sin (\beta \kappa
x)\Big]+O(b^{-4})  \label{eq12}
\end{equation}
where $\Omega =\kappa \,c\,(1+\varepsilon \,b^{-2})$. In the
Maxwellian limit ($b$ $\rightarrow $ $\infty $) $u(t,x)$
represents a stationary wave plus a uniform magnetostatic field
$B_{{\ell }}$. The boundary condition (\ref{eq10}) implies $\kappa
=n\pi /d$. The $\alpha $-term is a modulation of the magnetic
field caused by the stationary wave. By replacing this solution in
the Born-Infeld equation (\ref{eq11}), one obtains
\begin{equation}
\mathcal{BI\;}u(t,x)=-\frac{c^{2}\kappa E}{2b^{2}}\,\Big[\left(
E^{2}+2B_{{\ell }}^{2}+4\varepsilon \right) \cos (\kappa ct)\sin
(\kappa x)-2B_{{\ell }}\, E\left( \sin (2\kappa x)-\alpha \beta
^{2}\sin (\beta \kappa x)\right) \Big]+O(b^{-4}) \label{eq13}
\end{equation}
Therefore, in order that $u(t,x)$ in Eq.~(\ref{eq12}) be a
solution of Eq.~(\ref{eq11}) at the considered order, the values
of $\Omega $, $\alpha $ and $\beta $ should be
\begin{equation}
\Omega =\kappa c\left( 1-\frac{E^{2}+2B_{{\ell
}}^{2}}{4b^{2}}\right) \;, \hspace{0.2in}\alpha
=\frac{1}{4}\;,\hspace{0.2in}\beta =2\, .  \label{eq14}
\end{equation}
As said, a Lorentz boost along $z$\ will transform this field in a
transversal electric mode propagating in the waveguide. Since the
solution does not depend on $y$; the wave will result in a
TE$_{n0}$ mode. The Lorentz boost will not modify $x$ and $y$ in
Eq.~(\ref{eq7}), but $dt$ will be replaced by $\gamma
(V)(dt^{\prime }-Vc^{-2}dz^{\prime })$:
\begin{equation}
F\ =\ \frac{\partial u(t,x)}{\partial t}\ \gamma (V)\;dt^{\prime
}\wedge dy-\frac{\partial u(t,x)}{\partial t}\ \gamma
(V)Vc^{-2}dz^{\prime }\wedge dy+\ \frac{\partial u(t,x)}{\partial
x}\ dx\wedge dy \, . \label{eq15}
\end{equation}
Besides, the phase $\Omega t$ will change to $\Omega \gamma (V)(t^{\prime
}-Vc^{-2}z^{\prime })$, which means that the components of the wave vector
are
\begin{equation}
k_{t^{\prime }}=\omega =\Omega \gamma
(V)\;,\hspace{0.4in}k_{z^{\prime }}=\Omega \gamma (V)Vc^{-2}\; ,
\label{eq16}
\end{equation}
so the dispersion relation is
\begin{equation}
\omega ^{2}=k_{z^{\prime }}^{2}\,c^{2}+\Omega ^{2}  \label{eq17}
\end{equation}
The energy velocity $V$ can be recovered from Eqs.~(\ref{eq16})
and (\ref{eq17})) as
\begin{equation}
V=\frac{c^{2}k_{z^{\prime }}}{\omega }=\frac{\partial \omega
}{\partial k_{z^{\prime }}}\, .  \label{eq18}
\end{equation}
For a given frequency $\omega$, the wave number $k_{z^{\prime }}$
and the energy velocity $V$ depend on the wave amplitude $E$ and
the magnetostatic field $B_{{\ell }}$ through the functional form
of $\Omega$ (see Eqs.~(\ref{eq14}) and (\ref{eq17})). In
particular, Eq.~(\ref{eq17}) tells us that the minimum frequency
that propagates in the waveguide is
\begin{equation}
\omega _{cutoff} =\Omega _{min}=\Omega (n=1)=\frac{\pi
\,c}{d}\left( 1-\frac{E^{2}+2B_{{\ell }}^{2}}{4b^{2}}\right)
+O(b^{-4})  \label{eq19}
\end{equation}
The presence of $B_{{\ell }}$ in the cutoff frequency (a typical
nonlinear effect) offers a way for controlling the energy flux in
the guide. Namely, a given frequency $\omega $ could be larger
than $\omega _{cutoff}$ when the magnetostatic field $B_{{\ell }}$
is on, so the wave propagates. But the same $\omega $ could become
lower than $\omega _{cutoff}$ if $B_{{\ell }}$ is turned off.
Thus, one could allow the wave propagate or not by switching on
and off the magnetostatic field along the waveguide.

\section{The transmitted power}
We will calculate the energy flux in the waveguide. The energy
flux per unit of time and area along the $z$-direction is the
component $T_{t^{\prime }}^{\;\;z^{\prime }}$ of the
energy-momentum tensor. The non-diagonal components of $T^{\mu \nu
}$ in Born-Infeld electrodynamics are particularly simple (see for
instance Ref.$\,$\cite{Aiello}):
\begin{equation}
T_{t^{\prime }}^{\;\;z^{\prime }}=-\frac{1}{4\pi }\;F_{t^{\prime
}\mu }\; \mathcal{F}^{z^{\prime }\mu } \, . \label{eq20}
\end{equation}
In the case under study it is
\begin{equation}
T_{t^{\prime }}^{\;\;z^{\prime }}=-\frac{1}{4\pi
}\;\frac{F_{t^{\prime }\mu }\;F^{z^{\prime }\mu
}}{\sqrt{1+\frac{2S}{b^{2}}}}=-\frac{1}{4\pi }\;
\frac{F_{t^{\prime }y}\;F_{z^{\prime
}y}}{\sqrt{1+\frac{2S}{b^{2}}}}=\frac{\gamma (V)^{2}V}{4\pi
c^{2}}\;\frac{(\partial u/\partial
t)^{2}}{\sqrt{1+\frac{2S}{b^{2}}}}=\frac{\omega \ k_{z^{\prime
}}}{4\pi \,\Omega ^{2}}\ \frac{(\partial u/\partial
t)^{2}}{\sqrt{1+\frac{2S}{b^{2}}}} \label{eq21}
\end{equation}
Therefore, the time-averaged transmitted power in the waveguide is
\begin{equation}
{\cal{P}}=\int dx\ dy\ \langle T_{t^{\prime }}^{\;\;z^{\prime
}}\rangle =\frac{\omega \ k_{z^{\prime }}\ Area\ E^{2}}{16\ \pi \
\kappa ^{2}} \;\left( 1+\frac{3\ E^{2}}{32\
b^{2}}+O(b^{-4})\right) \, ,  \label{eq22}
\end{equation}
where $\langle\ \rangle$ means the integration in a period divided
by the period. If the Born-Infeld constant $b$ goes to infinity,
then the Maxwellian result for the transmitted power is recovered.
The Born-Infeld correction is expected to be very weak. However,
if one manages to operate the waveguide near the cutoff frequency,
then one could benefit from the fact that $\partial k_{z^{\prime
}}/\partial \omega $ diverges at $\omega _{cutoff}$. This implies
that the non-linear contributions to $k_{z^{\prime }}$ are
amplified near the cutoff frequency. In fact, according to
Eqs.~(\ref{eq17}) and (\ref{eq19}), $k_{z^{\prime }}$ is
\begin{equation}
k_{z^{\prime }}\ =\ c^{-1}\sqrt{\omega ^{2}-\Omega _{min}^{2}}\ =\
c^{-1}\sqrt{\omega ^{2}-\left( \frac{\pi \,c}{d}\right) ^{2}} +\
\frac{\pi ^{2}c}{4\ d^{2}b^{2}}\ \frac{E^{2}+2B_{{\ell
}}^{2}}{\sqrt{\omega ^{2} - \left( \frac{\pi \,c}{d}\right)
^{2}}}\ +\ O(b^{-4})\, .  \label{eq23}
\end{equation}
Thus, the slight change in the transmitted power along the
waveguide caused by switching on and off the longitudinal
magnetostatic field $B_{{\ell }}$ can be approximated as
\begin{equation}
\frac{\Delta \cal{P}}{\cal{P}}\ =\ \frac{\Delta k_{z^{\prime
}}}{k_{z^{\prime }}}\ =\ \frac{1}{2}\ \frac{B_{{\ell
}}^{2}/b^{2}}{\left( \frac{\omega \,d}{\pi \,c}\right) ^{2}-1}\ +\
O(b^{-4})\, , \label{eq24}
\end{equation}
the approximation being valid if the expression (\ref{eq24}) is
much smaller than $1$.

\section{Conclusion}
According to Eq.~(\ref{eq24}), by tuning the wave frequency
$\omega $ very close to the cutoff frequency $\omega
_{cutoff}\simeq \pi c/d$ one could improve the chance of revealing
the tiny nonlinear effect on the transmitted power and thus
determining the Born-Infeld constant $b$. This fine tuning could
be achieved by moving the power sensor in search of the frame
where the wave number is nearly zero (so the energy velocity $V$\
is nearly zero). The detection of the effect (\ref{eq24}) on the
transmitted power would constitute a clear manifestation of the
nonlinear behavior in the propagation of electromagnetic waves in
vacuum. Of course, since we have solved Born-Infeld theory at the
order $b^{-2}$, the obtained results are also valid for any
non-linear electrodynamics having the same form at that order:
$L=(1/4 \pi)\left[S-b^{-2}(S^2+P^2)/2 \right]+O(b^{-4})$. In
particular, since the pseudoscalar $P$ vanishes for the solution
(\ref{eq7}), our results are shared with the Euler-Heisenberg
Lagrangian, the weak-field limit for the one-loop approximation of
QED \cite{eh,schw,ditt},
\begin{equation}
L_{EH}=\frac{1}{4\, \pi}\left[S-4\,\mu\,\left(S^2+\frac{7}{4}\,
P^2\right)\right],\label{eq25}
\end{equation}
$\mu$ being
\begin{equation}
\mu=\frac{2\, \alpha^2}{45\, m_e c^2}\, \left(\frac{\hbar}{m_e
c}\right)^3,\label{eq26}
\end{equation}
where $\alpha$ is the fine-structure constant. In fact, those
solutions having $P=0$ are common to both theories provided that
the Born-Infeld constant $b$ is identified with $1/\sqrt{4\,
\mu}$, which has a magnitude of $10^{20} V/m$ \cite{Jackson}.

\end{document}